\documentclass[epj,final]{svjour}

\usepackage{graphicx}
\usepackage{amsmath}
\usepackage{amssymb}
\usepackage{amsfonts}
\usepackage{color}
\usepackage{bm}
\usepackage{times}

\usepackage{xypic} 

\newcommand{\eqname}[1]{\label{eq:#1}}
\newcommand{\bgar}{\begin{eqnarray}}
\newcommand{\enar}[1]{\label{eq:#1}\end{eqnarray}}

\newcommand{\vv}{ {\bf v}}

\newcommand{\eq}[1]{(\ref{eq:#1})}

\newcommand{\ahd}{\hat a^\dagger}
\newcommand{\ah}{\hat a}

\newcommand{\fhd}{\hat f^\dagger}

\newcommand{\upa}{\uparrow}
\newcommand{\doa}{\downarrow}

\begin{document}

\title{Algebraic Geometry tools for the study of entanglement: an application to spin squeezed states}

\authorrunning{A. Bernardi and I. Carusotto}
\titlerunning{Algebraic Geometry tools for the study of entanglement.}

\author{Alessandra Bernardi\inst{1} and Iacopo Carusotto\inst{2}}

\institute{GALAAD, INRIA M\'editerran\'ee, 2004 route del Lucioles, BP 93, F-06902 Sophia 
Antipolis, Cedex France. \\ \email{alessandra.bernardi@inria.fr} \and INO-CNR BEC Center and Dipartimento di Fisica, Universit\`a di Trento, via Sommarive 14, I-38123 Povo, Trento, Italy. \\ \email{carusott@science.unitn.it}}

\date{\today}

\abstract
{ 
A short review of Algebraic Geometry tools for the decomposition of tensors and polynomials is given from the point of view of applications to quantum and atomic physics.
Examples of application to assemblies of indistinguishable two-level bosonic atoms are discussed using modern formulations of the classical Sylvester's algorithm for the decomposition of homogeneous polynomials in two variables. In particular, the symmetric rank and symmetric border rank of spin squeezed states is calculated as well as their Schr\"odinger-cat-like decomposition as the sum of macroscopically different coherent spin states; Fock states provide an example of states for which the symmetric rank and the symmetric border rank are different.
%\keywords{\textbf{MSC2010}: 81Qxx -- 14Q05 -- 13P05}
}

% 05.30.Jp}

\maketitle

\section{Introduction}

The mathematical research on tensor decomposition is generally considered to have started out in the late XIX century with the pioneering work of J. J. Sylvester on the decomposition of a homogeneous polynomial  in two variables as the sum of powers of linear forms.
Still now, this kind of questions are attracting a strong interest from the algebraic geometry community and they have been related to the problem of decomposing a multi-dimensional, multi-index tensor as the sum of simpler, rank-1 objects; the minimum number of elements of such a  decomposition is called the {\em rank} of the tensor. Given the isomorphism between homogeneous polynomials and totally symmetric tensors, Sylvester's result is a first example in this program.
In the last years, the algebraic geometry tools for tensor decomposition are starting being applied in a number of application fields, ranging from
Signal Processing \cite{comon1,cm,bbcm},
%Antenna Array Processing 
%\cite{dm}, \cite{cafc} 
Telecommunications 
%\cite{vp}, \cite{chevalier}, 
\cite{sgb,lc}
to 
Phylogenetics \cite{ar,friedland,bo},
Chemometrics \cite{bro} 
%or Psychometrics \cite{kk}.
and 
Algebraic Statistics \cite{sz,os},
Geometric complexity theory \cite{ms2001,ms2008,bmlw},
Arithmetic complexity \cite{kruskal,bcrl,strassen,landsberg},
Data Analysis (e.g. Independent Component Analysis \cite{comon,cardoso,js}
%\cite{dh}
).

In physics, the Hilbert space of a composite quantum system is the tensor product of the Hilbert spaces of the constituent sub-systems, e.g. the individual particles forming a quantum gas or the entangled, yet physically separated sub-systems of quantum information processing devices.
Also in this context, it is often of interest to expand a given quantum state as the sum of factorizable states, i.e. tensor products of pure states of each sub-system. The minimum number of factorizable states that have to be summed to recover a given state  is called the {\em Schmidt rank} of the state and is often mentioned as a possible measure of entanglement~\cite{Schmidt-rank,Schmidt-rank2}. 
In many cases, the decomposition of the state has itself a physical interest, as it highlights an underlying Schr\"odinger cat structure as a linear superposition of macroscopically different states. 

In the present paper, we investigate the application of a algebraic geometry techniques to the calculation of the rank and of the decomposition of quantum states of actual physical interest of an assembly of $N$ indistinguishable two-level atoms, in particular spin squeezed states and Fock states.
On one hand, two level atoms are amenable to a description in terms of bivariate homogeneous polynomials that can be tackled using Sylvester's algorithm and its modern extensions \cite{cs,bgi,bcmt}.
On the other hand, spin squeezed states have been proposed as a practical tool to overcome the shot noise limit in precision interferometrical measurements~\cite{spin-squeezing-th,ueda,two-mode-BEC-th}: experimental demonstrations of this idea have been reported using either the internal atomic level structure in a cloud of thermal~\cite{spin-squeezing-exp} or Bose-condensed~\cite{spin-squeezing-exp-BEC} atoms or the atomic center-of-mass motion in a double well geometry~\cite{two-mode-BEC-exp}. 

The article is organized as follows: the relation between quantum states, tensors and polynomials is reviewed in Sec.\ref{sec:generals}. A general introduction to the concept of symmetric rank and symmetric border rank is given in Sec.\ref{sec:rank}. The basic definitions are given in Sec.\ref{sec:phys_int}, together with a physical interpretation in the second quantization formalism: for symmetric (skew-symmetric) tensors describing a Bose (Fermi) system, the decomposition of the physical state is made over Har\-tree (Hartree-Fock) states. A modern formulation of Syl\-ve\-ster's algorithm of \cite{bgi} is presented in Sec.\ref{sec:algo}.

Sec.\ref{sec:applications} is devoted to the application of these concepts to specific physical states, namely spin squeezed states and Fock states. The calculation of the symmetric rank and border rank of spin squeezed states is presented in Sec.\ref{sec:SSS}: the symmetric rank and symmetric border rank turn out to be always equal for this class of states, but interesting features are found whenever the squeezing parameter is a rational multiple of $2\pi$, $\mu= 2 \pi p/q$. For these values, the symmetric rank is lower than its generic value; as originally predicted in~\cite{indiani}, the state can be decomposed as the sum of $q$ macroscopically different coherent spin states pointing on equispaced directions along the horizontal plane orthogonal to the quantization axis.
Sec.\ref{sec:Fock} is devoted to Fock states with a well defined occupation number of the two states, for which the symmetric rank and symmetric border rank turn out to be different. Explicit decompositions are given as an exact sum or as the limit of sums of coherent spin states aligned on the horizontal plane.

Conclusions and perspectives are summarized in Sec.~\ref{sec:conclu}.

%In a system of many indistinguishable quantum particles, the indistinguishability postulate imposes that the wavefunction has to be either totally symmetric or totally anti-symmetric under particle exchange, depending on the bosonic (integer spin) or fermionic (half-integer spin) nature of the constituent particles.

\section{Quantum states, tensors and polynomials}
\label{sec:generals}

We consider a system of $N$ quantum particles. The Hilbert space $\mathcal{H}_1$ for each individual atom is assumed to have dimension $n$. The Hilbert space of the $N$-body system is obtained as the tensor product of $N$ copies of the $n$-dimensional single particle Hilbert space $\mathcal{H}_1$. In the case of indistinguishable bosonic (i.e. with integer spin) particles, the indistinguishability postulate of quantum mechanics imposes that only the totally symmetric states under particle exchange are physically relevant, which amounts to restricting our attention to the subspace $\mathcal{H}_s=S^N \mathcal{H}_1\subset \bigotimes^N \mathcal{H}_1$ of completely symmetric tensors. 

Using the $\{ e_1,\ldots ,e_n \}$ orthonormal basis of $\mathcal{H}_1$, a symmetric tensor $t$ can be written as 
\begin{equation}
t=\sum_{1\leq i_{j\in\{1\ldots N\}} \leq n} a_{i_1,\ldots ,i_N} e_{i_1}\otimes \ldots \otimes e_{i_N}
\eqname{tensor-t}
\end{equation}
where the coefficients tensor $a$ (i.e. the many-body wavefunction in the $\{ e_1,\ldots ,e_n \}$ basis) is symmetric under exchange, \\
$a_{\ldots i_j\ldots i_{j'}\ldots}=a_{\ldots i_{j'}\ldots i_{j}\ldots}$ for all $i_j,i_{j'}=1,\ldots,n$.

The standard isomorphism (i.e. biunivocal linear map) between symmetric tensors and homogeneous polynomials associates the tensor $t\in S^N\mathcal{H}_1$ defined in \eq{tensor-t} to the following homogeneous polynomial $p(x_1,\ldots,x_n)$ of degree $N$,
\begin{equation}
p(x_1,\ldots,x_N)=\sum_{1\leq i_{j\in\{1\ldots N\}} \leq n} a_{i_1,\ldots ,i_N} x_{i_1}\ldots x_{i_N}.
\eqname{poly-t}
\end{equation}

A related isomorphism between physical states and polynomials naturally arises within the second quantization description of many-body states in terms of the usual creation operators $\hat{a}_i^\dagger$ for an atom in the $i$ state. 
Each monomial $x_1^{n_1}\cdots x_N^{n_N}$ of degree $N$ is associated to a (unnormalized) physical state with a well defined number of particles $n_i$ in each single particle state $i=1,\ldots,n$ and total number of particles $\sum_{i=1}^d n_i=N$,
\begin{equation}
x_1^{n_1}\cdots x_N^{n_N}\;  \longleftrightarrow\; %\frac{1}{\sqrt{N!}}
(\hat{a}_1^{\dagger})^{n_1} \cdots (\ah_n^{\dagger})^{n_n} \vert \textrm{vac}\rangle.
\eqname{bos_2q} 
\end{equation}
It is immediate to see that this isomorphism can be extended by linearity to the whole space of polynomials and to the whole many-body Hilbert space $\mathcal{H}_s$. With the proper normalization factor, the isomorphism between polynomials and physical many-particle states induced in second quantization by \eq{bos_2q} is fully equivalent to the standard one of the mathematical literature summarized by \eq{tensor-t} and \eq{poly-t}.

An analogous isomorphism between physical states and Grass\-mann-generalized polynomials in anti-commuting variables can be defined for systems of fermionic (i.e. with half-integer spin) particles for which the many-body wavefunction is totally antisymmetric under exchange. Such states correspond to skew-symmetric tensors and, within second quantization, can be written in the form \eq{bos_2q} in terms of creation operators $\hat{a}_i^\dagger$ which now anti-commute.

\section{Rank and Border rank}
\label{sec:rank}

\subsection{Definition and physical interpretation}
\label{sec:phys_int}
For a given symmetric tensor $t\in S^N\mathcal{H}_1$, its {\em symmetric rank} $sr(t)$ is defined as the minimum integer $r$ for which there exists a set of (not necessarily normalized nor orthogonal) vectors $\vv_1, \ldots , \vv_r\in \mathcal{H}_1$ such that
\begin{equation}\eqname{t}
t=\sum_{i=1}^r  \vv_i^{\otimes N}.
\end{equation}
The {\em symmetric border rank} $sbr(t)$ of a symmetric tensor $t\in S^N\mathcal{H}_1$ is defined as the minimum integer $s$ for which there exists a sequence of symmetric tensors $t_i\in S^N\mathcal{H}_1$ of symmetric rank $s$ tending to $t$,  $\lim_{i\to \infty} t_i=t$. 
Obviously, one has
\begin{equation}
sbr(t)\leq sr(t).
\end{equation}
From a physical standpoint, the symmetric rank $sr$ has a simple interpretation in terms of the decomposition of a bosonic many-particle state as the sum of Hartree states where all $N$ atoms share the same single-particle state $\vv_i$. The symmetric rank $sr$ of a state $|\psi\rangle$ is then the minimum number of terms that have to appear in the sum in order to exactly reconstruct the state $|\psi\rangle$.

This decomposition can be straightforwardly reformulated within second quantization in terms of the associated polynomial. Given a homogeneous polynomial $p(x_1, \ldots , x_n)$ of degree $N$ in $n$ variables, the symmetric rank $sr(p)$ is the minimum integer $r$ for which there exists a decomposition as the sum of $r$ $N$-th powers of linear forms,
\begin{equation}\eqname{p}
%p(x_1, \ldots , x_n)=\sum_{i=1}^r l_i^{ N}
p(x_1, \ldots , x_n)=\sum_{i=1}^r \left(\sum_{j=1}^n \phi_{i,j} x_j\right)^N
\end{equation}
with $\phi_{i,j}\in \mathbb{C}$.

Identifying again the variables $x_1,\ldots,x_n$ of the polynomial with the creation operators $\ahd_i$ of the atom in the $i=1,\ldots,n$ single-particle state, the above definition of symmetric rank in terms of Hartree states is immediately recovered: it is the minimum integer $r$ for which 
\begin{equation}
|\psi\rangle=\sum_{i=1}^r \left(\sum_{j=1}^n \phi_{i,j} \ahd_j\right)^N |\textrm{vac}\rangle.
\end{equation}
The complex coefficients $\phi_{i,j}$ of the $i$-th linear form on the $j$-th variable of the polynomial have the physical interpretation of the $j$-state component of the $i$-th Hartree wavefunction.

Analogously the symmetric border rank $sbr(p)$ of a homogeneous polynomial $p(x_1, \ldots , x_n)$ of degree $N$ in $n$ variables is defined as the minimum integer $s$ for which there exists a sequence of homogeneous polynomials  $p_i(x_1, \ldots , x_n)$ of degree $N$ in $n$ variables of symmetric rank $s$ tending to $p(x_1, \ldots , x_n)$,  $\lim_{i\to \infty} p_i(x_1, \ldots , x_n)=p(x_1, \ldots , x_n)$.

These definitions can be generalized to not necessarily symmetric tensors, e.g. the ones describing the state of distinguishable quantum particles. For a given  tensor $t\in \mathcal{H}_1^{\otimes N}$, its {\em rank} $r(t)$ is defined as the minimum integer $r$ for which there exists a family of vectors $\vv^{(j)}_{i}\in \mathcal{H}_1$ with $j=1, \ldots , N$ and $i=1, \ldots , r$, such that
\begin{equation}
t=\sum_{i=1}^r  \vv^{(1)}_{i}\otimes \cdots \otimes \vv^{(N)}_{i}.
\end{equation}
The {\em border rank} $br(t)$ of a  tensor $t\in \mathcal{H}_1^{\otimes N}$ is defined as the minimum integer $s$ for which there exists a sequence of  rank-$s$  tensors $t_i\in  \mathcal{H}_1^{\otimes N}$ such that $\lim_{i\to \infty} t_i=t$.
Note that in the absence of the symmetry requirement, the single-particle wavefunctions $\vv^{(1)}_{i},\ldots, \vv^{(N)}_{i}$ of the different $N$ particles are no longer necessarily equal.
The so-called {\em Comon's conjecture}~\cite{bglm} states that if $t\in S^N\mathcal{H}_1$, then $sr(t)=r(t)$. Such a conjecture is proved to be true only in few cases~\cite{bglm,bb}. An analogous conjecture can be formulated in terms of the border rank and the symmetric border rank (see e.g.~\cite{bgl}).

Analogous concepts can be introduced for skew-symmetric tensors $t\in \bigwedge^N\mathcal{H}_1$. In this case, the {\em skew-symmetric rank} $ar(t)$ is defined as the minimum integer $r$ for which there exists a family of orbitals $\vv^{(j)}_{i}\in \mathcal{H}_1$ with $j=1, \ldots , N$ and $i=1, \ldots , r$, such that
\begin{equation}
t=\sum_{i=1}^r  \vv^{(1)}_{i}\wedge \cdots \wedge \vv^{(N)}_{i}.
\eqname{skew_decomp}
\end{equation}
The {\em skew-symmetric border rank} $abr(t)$ of a skew-symmetric tensor $t\in \bigwedge^N\mathcal{H}_1$ is defined~\footnote{
It is interesting to note that the rank of a general skew-symmetric $t\in \bigwedge^2\mathcal{H}_1$ is equal to half the rank of the skew-symmetric matrix $T$ that is obtained by writing $t$ in an arbitrary basis of $\mathcal{H}_1$~\cite{harris}.} as the minimum integer $s$ for which there exists a sequence of skew-symmetric rank $s$ skew-symmetric tensors $t_i\in  \bigwedge^N\mathcal{H}_1$ such that $\lim_{i\to \infty} t_i=t$.

In physical terms, the decomposition \eq{skew_decomp} corresponds to expanding the many-particle state of a Fermi system as the sum of Hartree-Fock states,
\begin{equation}
|\psi\rangle=\sum_{i=1}^r \left[\prod_{k=1}^N \left( \sum_{j=1}^n \phi^{(k)}_{i,j} \fhd_j \right)\right] |\textrm{vac}\rangle,
\end{equation}
where, $\phi^{(k)}_{i,j}$ is the  $j$-state component of the $k$-th orbital $k=1,\ldots,N$) of the $i$-th Hartree-Fock state ($i=1,\ldots,r$) and $\fhd_j$ are the (fermionic) creation operators for a particle in the $j$ single-particle state; as usual for Fermi systems, creation operators satisfy the anti-commutation rules of Grassmann variables,
\begin{equation}
\fhd_i \fhd_j -\fhd_j\fhd_i =0
\end{equation}

\subsection{Algorithm}
\label{sec:algo}

Let now $\mathcal{H}_1$ be a 2-dimensional vector space defined over the complex numbers.
The algorithm to compute the symmetric rank of a homogeneous polynomial $p$ in two variables (i.e. the symmetric rank of a symmetric tensor $t\in S^{N}\mathcal{H}_1$) is classically attributed to J. J. Sylvester. A more modern proof of it is presented in~\cite{cs}. We actually  refer to a simplified version of the same available in~\cite{bgi}. From now on we will interchangeable use the polynomial and symmetric tensor language.
\\
Before going into the details of the algorithm we need to define the so-called {\em Catalecticant matrices}~\cite{geremita,kanev} $C_{N-r,r}(p)$ ($C_{N-r,r}(t)$) associated to a given bivariate polynomial $p(x,y)$ (to a given tensor $t\in S^{N}\mathcal{H}_1$),
\begin{equation} 
p(x,y)=\sum_{i=0}^{N} {N\choose i}\,a_{i}x^{i}y^{N-i}.
\end{equation}  
%and $t =(b_{i_1},...,b_{i_d})_{i_{j}=0,1;\; j=1, \ldots , d}\in S^{d}V$ be the symmetric tensor associated to $p$, as we did at the beginning of section. 
The $r$-th  Catalecticant matrix $C_{N-r,r}(p)=(c_{k,j})$ associated
to $p$ is the $(N-r+1) \times (r+1)$ matrix with
entries:
$c_{k,j}=a_{k+j-2}$ with $k=1, \ldots , N-r+1$ and $j=1, \ldots , r+1$. 

It is classically known~\cite{kanev} that, for a given homogeneous bivariate polynomial $p(x,y)$ of degree $N$, if $r$ is the minimum integer for which all the minors of $C_{N-r,r}(p)$ of order $(r+1)$ vanish then $sbr(p)=r$. Moreover in~\cite{cs} and in~\cite{bgi} it is shown that if $sbr(p)=r$, then $sr(p)$ can only be equal either to $r$ or to $N-r+1$. The way to distinguish among these last two possibilities is to verify if a certain polynomial of degree $r$ constructed via elements in the kernel of $C_{N-r,r}(p)$ (see precisely below in the algorithm) has or not distinct roots. So the Algorithm that we are going to describe works as follows: for a given $p(x,y)$ find firstly its symmetric border rank, then compute its symmetric rank, and, finally, give the decomposition in the form \eq{p}.

It is worth to remark here that, over the complex numbers, the symmetric border rank of any homogeneous $p(x,y)$ is always bound from above by the symmetric rank of a generic bivariate polynomial, that is,
\begin{equation}
sbr(p)\leq \left\lceil \frac{N+1}{2} \right\rceil.
\end{equation}

Moreover, since if $sbr(p)=r$ then $sr(p)\in \{r,N-r+2\}$, we easily deduce that if $p(x,y)$ is such that its symmetric border rank $sbr(p)$ is strictly less than both its symmetric rank $sr(p)$ and the generic symmetric rank  $\left\lceil \frac{N+1}{2} \right\rceil$, then the symmetric rank of $p(x,y)$ is strictly bigger then the generic value $\left\lceil \frac{N+1}{2} \right\rceil$.

We are now ready to state the Algorithm that for a given homogeneous bivariate polynomial of degree $N$, calculates its symmetric border rank, its  symmetric rank and its decomposition as the sum of $r$ $N$-th powers of linear forms as in \eq{p}.
\\
\rule{8,8cm}{0.1mm}
\\
\textbf{Algorithm}
\\
\rule{8,8cm}{0.1mm}
\\
\textbf{Input}: The projective class of a degree $N$ homogeneous polynomial $p(x,y)$ in two variables. 
\\
\rule{8,8cm}{0.1mm}
\\
\textbf{Output 1}: $sbr(p)$.\\
\textbf{Output 2}: $sr(p)$.\\
\textbf{Output 3}: a decomposition of $p$ as in \eq{p}.\\
\\
\rule{8,8cm}{0.1mm}
\begin{enumerate}
\item Initialize $r=0$;
\item\label{oursyl2} Increment $r\leftarrow r+1$;
\item Compute  $C_{N-r,r}(p)$'s $(r+1)\times (r+1)$-minors; 
\\ if they
are not all equal to zero then go to step \ref{oursyl2} (note that this can only happen if $r\leq \left\lceil\frac{N+1}{2}\right\rceil$);
\\ else, \\
$\hbox{\textbf{Output 1}: } sbr(p)=r.$
Go to step \ref{oursyl4}.
\item\label{oursyl4} Choose a solution $(\overline{u}_{0}, \ldots ,
\overline{u}_{r})$ of the system 
\begin{equation}
C_{N-r,r}(p)\cdot (u_0,\ldots,u_r)^{t} =0.
\end{equation}
If the polynomial
\begin{equation}\eqname{q}
q(t_0,t_1)=\overline{u}_0t_0^r+\overline{u}_1t_0^{r-1}t_1+\cdots+\overline{u}_rt_1^r
\end{equation}
has distinct roots, 
\begin{equation}
q(t_0,t_1)=\mu \prod_{k=1}^{r} (\beta_{k} t_{0}-\alpha_{k} t_{1}),
\end{equation}
then 
\\
$\hbox{\textbf{Output 2}: } sr (p) = r$
and go to step \ref{oursyl6}, 
\\
otherwise \\
$\hbox{\textbf{Output 2}: } sr (p) = N-r+2$
and go to step  \ref{oursyl5}.
\item\label{oursyl5} Choose a solution $(\overline{u}_{0}, \ldots ,
\overline{u}_{sr(p)})$ of the system 
\begin{equation}
C_{N-sr(t),sr(t)}(p)\cdot (u_0,\ldots,u_{sr(p)})^{t} =0.
\end{equation}
The polynomial
\begin{equation}
q(t_0,t_1)=\overline{u}_0t_0^{sr(p)}+\overline{u}_1t_0^{{sr(p)}-1}t_1+\cdots+\overline{u}_{sr(p)}t_1^{sr(p)}
\end{equation}
has distinct roots, 
\begin{equation}
q(t_0,t_1)=\mu \prod_{k=1}^{sr(p)} (\beta_{k} t_{0}-\alpha_{k} t_{1}).
\end{equation}
Go to step \ref{oursyl6}.
\item\label{oursyl6} Solve the linear system in the variables $\lambda_1, \ldots ,\lambda_{sr(p)}$ given by 
\begin{equation}\eqname{decomposition}
p(x,y)=\sum_{k=1}^{sr(p)} \lambda_{k} (\alpha_{k} x + \beta_{k} y)^{N}.
\end{equation}
For such $\lambda_i$'s, the decomposition \eq{decomposition} is our \textbf{Output 3}.
\end{enumerate}
\rule{8,8cm}{0.1mm}
\\
\\
Observe that, for the polynomials  $p(x,y)$ for which $sbr(p)<sr(p)$, the polynomial $q(t_0,t_1)$ defined in \eq{q} has not distinct roots, i.e. there exist integers $s_1, \ldots , s_{r'}$ with $r'<sr(p)$ and $\sum_{k=1}^{r'}s_k=sbr(p)<sr(p)$  such that the polynomial $q(t_0,t_1)$ obtained as in step \ref{oursyl4} of our algorithm can be written as follows,
\begin{equation}\eqname{qnotdistinct}
q(t_0,t_1)=\mu \prod_{k=1}^{r'} (\tilde\beta_{k} x_{0}-\tilde\alpha_{k} x_{1})^{s_k}
\end{equation}
We like to stress here that those roots  have an algebraic meaning even if they are not distinct (see~\cite{bgi}). In fact, in this case, there exist $\tilde\lambda_1, \ldots , \tilde\lambda_{r'}\in \mathbb{C}$ and polynomials in two variables $F_1, \ldots , F_{r'}$ of degrees $(s_1-1), \ldots , (s_{r'}-1)$ respectively   such that  
\begin{equation}
p(x,y)= \sum_{k=1}^{r'}\tilde\lambda_k (\tilde\alpha x + \tilde\beta y )^{N- (s_k-1)}F_k.
\end{equation}

Before moving to the physical applications, we wish to stress that in the present section we have chosen to restrict our attention to the algorithm that compute the symmetric rank  for a polynomial in two variables because this is the relevant case for the applications that we are going to discuss in the next section. A more general algorithm that computes the symmetric rank of a homogeneous polynomials in $n\geq 2$ variables can be found in~\cite{bcmt}.

\section{Examples of application}
\label{sec:applications}

We now proceed to illustrate the application of the general concepts discussed in the previous sections to specific examples of current interest in the field of atomic physics. In particular, we focus our attention on the $n=2$ case where each of the $N$ bosonic atoms has only two available states, denoted as $\upa$ and $\doa$ respectively: in this case, the symmetric border rank, the symmetric rank and the decomposition can be obtained using the algorithm reviewed in Sec.\ref{sec:algo}.

\subsection{Spin Squeezed States}
\label{sec:SSS}

In the last years, {\em spin squeezed states} (SSS) have attracted quite some attention as a tool to improve the sensitivity of interferometric device for precision measurements, as they allow to overcome the so-called standard quantum limit imposed by shot noise~\cite{spin-squeezing-th,ueda}. Experimental demonstrations using different physical systems have been reported in~\cite{spin-squeezing-exp,spin-squeezing-exp-BEC,two-mode-BEC-exp}.

To generate a spin squeezed state, one can start from a {\em coherent spin state} (CSS), i.e. a symmetric rank-1 Hartree state where all atoms share the same wavefunction. In the present case where the single atom Hilbert space is bidimensional, Har\-tree states can be parametrized in terms of a macroscopic spin vector moving on the surface of the so-called Bloch sphere. Its direction is specified by two angles $\theta,\varphi$ with $0\leq \theta \leq \pi$ and $0\leq \varphi < 2\pi$,
\begin{equation}
\eqname{CSS}
|\textrm{coh}:\theta,\varphi\rangle=\frac{1}{\sqrt{N!}}\,\left[\cos({\theta}/{2})\, \ahd_\upa + e^{i\varphi} \sin (\theta/{2})\, \ahd_\doa \right]^N\,|\textrm{vac}\rangle:
\end{equation}
the $\theta=0$ ($\theta=\pi$) case with the macroscopic spin pointing along (opposite to) the quantization axis $z$ corresponds to all atoms being in the $\upa$ ($\doa$) state. A macroscopic spin along the ``horizontal" plane orthogonal to the quantization axis corresponds to equally populated $\upa$ and $\doa$ states; the phase $\varphi$ corresponds to the angle made by the macroscopic spin and the $x$ axis.

As proposed in~\cite{ueda}, SSS can be obtained from a $\theta=\pi/2$, $\varphi=0$ CSS by means of a one-axis twisting of this state via the self-phase modulation Hamiltonian,
\begin{equation}
|SSS:\mu\rangle= \exp(-i\mu S_z^2/2)\,|\textrm{coh}:\theta=\pi/2,\varphi=0\rangle.
\eqname{SSS} 
\end{equation}
In elementary quantum mechanics, the total spin operator $S_z$ is defined as the sum of the corresponding $S_z^{(i)}$ operators acting on the $i$-th atom and leaving the other atoms unperturbed; the $\upa$ and $\doa$ states of the $i$-th atom are eigenstates of $S_z^{(i)}$ with eigenvalue $\pm 1/2$. In a second quantization formalism, the total spin operator $S_z$ has the simple form
\begin{equation}
S_z=\frac{1}{2}\left( \ahd_\upa \ah_\upa - \ahd_\doa \ah_\doa \right).
\end{equation}

The first step in view of computing the rank of the $|SSS\rangle$ state is to write it in a second quantization form as a polynomial in the creation operators $\ahd_{\upa,\doa}$. Expanding the coherent spin state \eq{CSS} and then applying the one-axis twist operators as in \eq{SSS}, one easily obtains the form
\begin{multline}
|SSS:\mu\rangle=\frac{1}{\sqrt{2^N\,N!}}\times \\ \times\sum_{k=0}^N {N \choose k}\,e^{-i\mu(k-N/2)^2/2}\,(\ah_\upa)^{N-k}\,(\ah_\doa)^k\,|\textrm{vac}\rangle.
\end{multline}
The key element of the algorithm discussed in Sec.\ref{sec:algo} are the catalecticant matrices $C_{N-r,r}(SSS)$ associated to the polynomial: in our case, their entries for $1\leq k\leq N-r+1$ and $1\leq j \leq r+1$ are equal to
\begin{equation}
c_{k,j}=\frac{1}{\sqrt{2^N\,N!}}\,\exp\left[-i\frac{\mu}{2}\left(k+j-2-\frac{N}{2}\right)^2\right].
\end{equation}
To determine whether the SSS has symmetric border rank $r$ one has to test the $r+1$-minors of the $r$-th catalecticant matrix $C_{N-r,r}(SSS)$.

Let $0\leq j_1,\ldots,j_{r+1}\leq (N-r+1)$ be the row indices of the rows that are involved in a given minor. Straightforward algebraic manipulation shows that the minor is zero if and only if the Vandermonde determinant~\cite{lang}
\begin{equation}
\begin{vmatrix}
1 & e^{-i\mu j_1} & \cdots & e^{-i\mu j_1 r} \\
1 & e^{-i\mu j_2} & \cdots & e^{-i\mu j_2 r} \\
\vdots  & \vdots  & \ddots & \vdots  \\
1 & e^{-i\mu j_{r+1}} & \cdots & e^{-i\mu j_{r+1} r}
\end{vmatrix}
=0
%\left| 
%\begin{array}{cccc}
%1 & e^{-i\mu j_1} & ldots & e^{-i\mu j_1 r \\
%\vdots 
%& \lambda - e & -f \\
%1 & -h & \lambda - i \end{array} \right|.\] 
\end{equation}
vanishes, i.e. if and only if
\begin{equation}
\prod_{1\leq j_\alpha < j_\beta \leq (r+1)}\left(e^{-i\mu j_\beta}-e^{-i\mu j_\alpha}\right)=0.
\end{equation}
In order for this equality to hold, one has to have $\mu (j_\beta-j_\alpha)\equiv 0\,(\textrm{mod}\,2\pi)$ for at least one pair $j_\alpha\neq j_\beta$.

As discussed in Sec.\ref{sec:algo}, the state has symmetric border rank equal to $r$ if and only if the rank of the $r$-th  catalecticant matrix $C_{N-r,r}(SSS)$ is not maximal (i.e. smaller than $r+1$) and the rank of all the previous $r'$-th  catalecticant matrices $C_{N-r',r'}(SSS)$ with $r'<r$ was maximal (i.e. equal to $r'+1$). Assuming that this latter condition is satisfied, the symmetric border rank is $r$ if for all possible choices for $j_1,\ldots,j_{r+1}$, there exists at least one pair $j_\alpha \neq j_\beta$ such that 
\begin{equation}
\mu (j_\beta-j_\alpha)\equiv 0\,(\textrm{mod}\,2\pi).
\end{equation}

If $\mu$ is an irrational multiple of $2\pi$, this condition can never be satisfied, so the symmetric border rank of the spin squeezed state  has the generic value $g_N=\left\lceil\frac{N+1}{2}\right\rceil$ that corresponds to have maximal symmetric border rank. 

If $\mu$ is a rational multiple of $2\pi$, i.e. $\mu=2\pi p/q$ ($p,q$ are coprime integers) and $q\leq r$, the pigeon hole principle guarantees that for any choice of row indices $0\leq j_1,\ldots,j_{r+1}\leq (N-r+1)$, there always exists a pair $j_\alpha\neq j_\beta$ such that $\mu(j_\alpha - j_\beta)\equiv0\,(\textrm{mod}\,2\pi)$. On the other hand, for $r< q$ there exists a set $j_1=1,\ldots,j_{r+1}=r+1$ such that all pairs $\mu(j_\alpha -j_\beta) \not\equiv0\,(\textrm{mod}\,2\pi)$.
Combining these two arguments into the algorithm of Sec.\ref{sec:algo}, we conclude that the symmetric border rank of the spin squeezed state with squeezing parameter $\mu=2\pi p/q$ is equal to $q$ if $q\leq g_N$ and equal to $g_N$ if $q\geq g_N$.

\begin{figure}[htbp]
\begin{center}
\includegraphics[width=0.95\columnwidth,angle=0,clip]{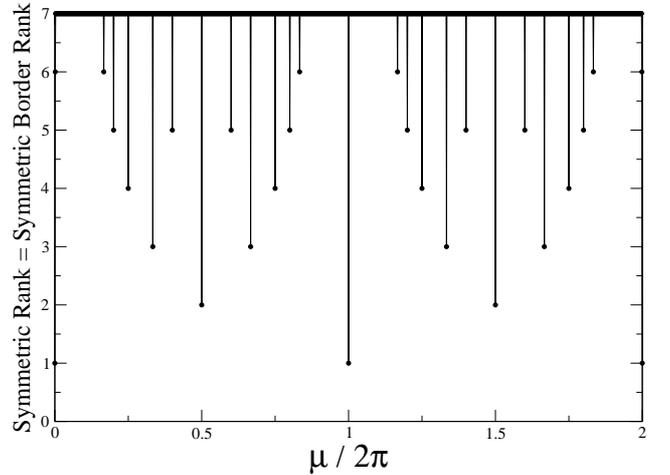}
\caption{
Numerical plot of the symmetric rank and symmetric border rank of spin squeezed states of the form \eq{SSS} as a function of the squeezing parameter $\mu$. Total number of particles $N=13$.
}
\label{fig:SSS_rank}
\end{center}
\end{figure}

In order to determine the value of the symmetric rank, we have to consider the  polynomial $q(t_0,t_1)$ defined in \eq{q} that is associated to vectors in the null space of $C_{N-r,r}(SSS)$  for $r=sbr(SSS)$

For $\mu=2\pi p/q$ and $q\leq g_N$, direct inspection shows that the $(r+1)$-component ($r=q$) vector
\begin{equation}
\left(1,0,\ldots,0,-e^{i\pi pq}e^{-i\pi p N}\right)^t
\end{equation}
always belongs to the null space of the $r$-th  catalecticant matrix $C_{N-r,r}(SSS)$, so that the $r$ roots of the associated polynomial
\begin{equation}
q(t_0,t_1)=t_0^{r}-e^{i\pi pq}e^{-i\pi p N} t_1^{r}
\eqname{qSSS}
\end{equation}
are always distinct. This shows that the symmetric rank is always equal to the symmetric border rank:
\begin{equation}
sr(SSS)=sbr(SSS).
\end{equation}

The situation is more delicate in the general case when $\mu$ is an irrational multiple of $2\pi$ or $\mu=2\pi p/q$ with $q> g_N$, so that the symmetric border rank has the maximal, generic value $sbr=g_N$. If $N$ is even, the only possibility for the symmetric rank is to be $sr=sbr=g_N$. If $N$ is odd, one has to assess whether $sr=g_N$ (the generic case) or $sr=g_N+1$. To this purpose, one has to repeat the procedure of calculation the roots of the polynomial associated with the null space of the $r=g_N+1$ catalecticant matrix. By means of a {\sc matlab/octave} code implementing the Algorithm described in Sec.\ref{sec:algo}, we have checked that the roots are always distinct, so that $sr=sbr=g_N$. 
These results on dependence of the symmetric rank of SSS on the squeezing parameter $\mu$ are summarized in Fig.\ref{fig:SSS_rank}.

Let us now move to the decomposition. For $\mu=2\pi/q$, the roots of the polynomial \eq{qSSS} are complex numbers of magnitude $1$ and equidistant phases by $2\pi/q$: this remarkable fact recovers a result originally found in~\cite{indiani} that the spin squee\-zed state with $\mu=2\pi p/q$ can be written as a Schr\"odinger-cat-like linear superposition of $q$ macroscopically different coherent spin states of the form \eq{CSS} with $\theta=\pi/2$ and equidistant phases $\phi=\phi_0+2\pi j/q$, $j=0,\ldots,q-1$; the initial phase is $\phi_0=0$ unless both $p$ and $q+N$ are odd, in which case is $\phi_0=\pi/q$.

The direction of the macroscopic spin is therefore always oriented along the plane orthogonal to the quantization axis and makes equidistant angles $\phi$ with the $x$ direction. A numerical solution of the linear system at the step 6 of the algorithm presented in Sec.\ref{sec:algo} shows that the weigths $\lambda_i$ of the $q$ coherent spin states have equal magnitude. These facts are summarized by the remarkable writing of the spin squeezed state in the form
\begin{multline}
|SSS:\mu=2\pi p/q\rangle= \\ =\sum_{j=1}^q\,\frac{e^{i\zeta_j}}{\sqrt{q}}\,|\textrm{coh}:\theta=\pi/2,\varphi=\phi_0+2\pi j/q\rangle
\end{multline}
with suitably chosen phases $\zeta_j$'2~\cite{indiani}. 

\subsection{Fock States}

\label{sec:Fock}

In the previous subsection we have seen that the symmetric rank and symmetric border rank are equal for spin squeezed states for all values of $\mu$. Here we illustrate a class of physically relevant states for which the symmetric rank and symmetric border rank are different. 
Having this distinction clear in mind is essential if the concept of Schmidt rank is to be used as a measure of entanglement~\cite{Schmidt-rank}.

In the mathematical literature, it is a well known fact that the symmetric rank and the symmetric border rank of bivariate monomials usually have different values: straightforward inspection of the form of the catalecticant matrices and of the associated polynomial \eq{q} shows that the symmetric rank of a monomial of the form $x^{N-k} y^k$, with $k\leq N/2$, is equal to $sbr=k+1$, while the symmetric rank is equal to $sr=N-k+1\geq sbr$. 

The physical states corresponding to monomials are Fock states (FS)
\begin{equation}
|\psi\rangle= (\ah_\upa)^{\dagger(N-k)} \, (\ah_\doa)^{\dagger k}\,|\textrm{vac}\rangle,
\end{equation}
where exactly $k$ particles sit in the $\doa$ state and the remaining $N-k$ particles sit in the $\upa$ state. The results on the symmetric rank and the symmetric border rank of monomials quoted above imply that a Fock state can be decomposed either as the exact sum of $sr=N-k+1$ CSS states or as the limit of the sum of a smaller (equal only if $k=N/2$ and $N$ even) number $sbr=k+1\leq sr$ of CSS states. 

The decomposition as an exact sum of $sr$ CSS states has the form
\begin{multline}
(\ah_\upa)^{\dagger N-k} (\ah_\doa)^{\dagger k}\,|\textrm{vac}\rangle=\frac{1}{(N-k+1){N \choose k}}\times \\
\times \sum_{j=0}^{N-k} \left( \ahd_\upa\,e^{-i \frac{2\pi k j }{(N-k+1)k} } +  \ahd_\doa\, e^{i \frac{2 \pi j}{N-K+1} (1-\frac{k}{N}) }   \right)^N\,|\textrm{vac}\rangle,
\eqname{sr_exp}
\end{multline}
where the relative phases of the $\upa,\doa$ amplitudes in the $sr$ CSS states have been carefully chosen in a way to have a complete destructive interference on all monomials $\ah_\upa^{N-k'}\ah_\doa^{\dagger k'}$ with $k'\neq k$.

The decomposition as the $\epsilon \to 0$ limit of a sum of $sbr$ CSS states has a similar structure:
\begin{multline}
(\ahd_\upa)^{N-k} (\ahd_\doa)^k\,|\textrm{vac}\rangle= \frac{1}{k {n \choose k}} \times \\
\times \lim_{\epsilon \to 0} \frac{1}{\epsilon^k}  \left[ \sum_{j=0}^k \left(  \ahd_\upa  + \epsilon\,e^{i \frac{2\pi}{k} j}   \ahd_\doa   \right)^N   -k\,  \ah_\upa^{\dagger N}  \right] \,|\textrm{vac}\rangle.
\eqname{sbr_exp}
\end{multline}
In contrast to \eq{sr_exp}, only the monomials  $\ah_\doa^{\dagger k'}$ with $k'<k$ vanish by destructive interference, while the ones with $k'>k$ disappear when the $\epsilon \to 0$ limit is taken.

\section{Conclusions and perspectives}
\label{sec:conclu}

In this paper we have reported an example of application of algebraic geometry tools to a simple model of quantum physics. We have calculated the symmetric rank and the decomposition of spin squeezed states and Fock states of an assembly of $N$ indistinguishable two-level bosonic atoms: the rank of spin squeezed states shows interesting features when the squeezing parameter has a rational value (in suitable units) so that the state can be decomposed as the superposition of a few macroscopically different coherent spin states. Fock states have the peculiar property that their symmetric rank and symmetric border rank have different values. 

A most challenging future perspective is to generalize the decomposition algorithms to the case of a large number $N$ of atoms and a large number $m$ of available states and then combine them with a numerical simulation of the time-evolution of an interacting system. First attempts in this direction used the semiclassical representations of a quantum field~\cite{quantumnoise} or a reformulation of the $N$-body problem in terms of stochastic Hartree~\cite{CCD} (Hartree-Fock~\cite{JC}) wavefunctions for systems of bosonic (fermionic) particles, but in many practical applications to strongly correlated many body systems this method lead to numerically untractable problems.

From a mathematical point of view, the computation of the symmetric border rank of a homogeneous polynomial of degree $N$ in $n>2$ variables (i.e. $N$ bosonic atoms in $n>2$ available states) is a still open problem as the equations defining the corresponding varieties are not yet understood (see e.g.~\cite{lo} and references therein). Nevertheless, algorithms to compute the symmetric rank and the decomposition have been recently developed~\cite{bcmt} using a generalized version of the catalecticant matrices, the so-called Henkel matrices. Their application to quantum physics problems is a natural next step for our research.

The situation for the skew-symmetric tensors describing systems of $N$ fermions is somehow specular: a criterion to determine the skew-symmetric border rank of a given skew-symmetric tensor of any $N$ was recently developed~\cite{lo2}. On the other hand, we are not aware of any algorithm to compute the skew-symmetric rank and the actual decomposition except for the first meaningful case $N=2$ where the tensor is a matrix.

\section*{Acknowledgments}

AB was partially supported by  Project Galaad of INRIA So\-phia Antipolis M\'editerran\'ee 
(France)  and   Marie Curie Intra-European Fellowships for Career Development (FP7-PEOPLE-2009-IEF): ``DECONSTRUCT". IC is grateful to Ch. Miniatura and P. Vignolo for the kind hospitality at INLN and acknowledges financial support from the ERC via the QGBE grant. 
Stimulating discussions with P. Hyllus are warmly ac\-know\-ledged.

\end{document}